\documentclass[runningheads,fleqn]{cl2emult}
\usepackage{amsmath}
\usepackage{makeidx}  
\usepackage{graphicx} 
\usepackage{subeqnar} 
\usepackage{multicol} 
\usepackage{cropmark} 
\usepackage{phys}     
\makeindex            

\usepackage{amsmath}   
\mathindent\parindent 
\def\I{{\cal I}} \def\K{{\cal K}} \def\Y{{\cal Y}} \def\J{{\cal J}}
\def\D{{\cal D}} \def\C{{\cal C}} \def\S{{\cal S}} 
\def\RU{{\rm U}}\def\RS{{\rm S}}
\def\bvc#1{\vec{#1}}
\def\boldtau{\vec{\tau}}
\def\sminus{{\textstyle -}} \def\splus{{\textstyle +}}
\def\Pr{^{\prime}}
\def\period{\ \ \ .} \def\comma{\ \ \ ,}
\def\bra#1{\big\langle #1 \big |} \def\ket#1{\big | #1 \big\rangle}
\def\braket#1#2{\big\langle #1 \big | #2 \big\rangle}
\def\Conjg#1{{#1}^{\ast}}

\def\gauntcof#1#2#3#4#5#6{
        {\cal G}\bigl(\thinspace
                #1,\thinspace
                #2;\thinspace
                #3,\thinspace
                #4;\thinspace
                #5,\thinspace
                #6\thinspace
        \bigr)
}
\def\Trace{{\rm T\kern-1.5pt r\kern 1.5pt}}
\def\Prpow#1{^{\prime\kern 1pt {#1}}}
\def\Fac{\kern-.2pt !}
\def\Dblfac{!\hskip-0.10em!}
\begin{document}

%
\title*{Full-Potential LMTO \protect\newline Total Energy and Force 
Calculations}
%
%
\toctitle{Full-Potential LMTO \protect\newline Total Energy and Force 
Calculations}
%
%
\titlerunning{Full-Potential LMTO}
%
\author{John M Wills\inst{1}
\and Olle Eriksson\inst{2}
\and Mebarek Alouani\inst{3}
\and David L. Price\inst{4}}
%
\authorrunning{J M Wills et al.} 

%
%
\institute{Los Alamos National Laboratory, Los Alamos, NM 87545, USA
\and Uppsala University, Uppsala, Sweden
\and IPCMS, 23 rue du Loess, 67037 Strasbourg, France
\and University of Memphis, Memphis, TN 38152, USA
}

\maketitle              

\begin{abstract}
The essential features of
a full potential electronic structure method using
Linear Muffin-Tin Orbitals (LMTOs)
are presented.
The electron density and potential in the this method are represented
with no inherent geometrical approximation.
This method allows the calculation of total energies and forces with 
arbitrary accuracy while sacrificing much of the efficiency and physical
content of approximate methods such as the LMTO-ASA method.
\end{abstract}

\section{Introduction}
This paper describes a particular implementation of a
full-potential electronic structure method using
Linear Muffin-Tin Orbitals (LMTO's) \cite{andersen,skriver}
as basis functions.
There have been several ``FP-LMTO'' implementations 
\cite{wills,springborg,methfessel,weyrich,savrasov}.
The one described here has not been published in detail,
although calculations performed with this method have been reported
for quite some time.\cite{wills}
There are many aspects to an electronic structure method.
This paper is focussed on those aspects which enable
a full potential treatment.
Relatively small details pertaining to full-potential methods
will be discussed while larger details having to do with, for example,
relativity will not be.

The emphasis of a variational 
full-potential method is somewhat different from
that of a method such as the LMTO-ASA method.  
The emphasis of the former is on the completeness of the basis 
while in the latter it is in the physical
content (and interpretability) of the basis.  
These concepts are, of course, intimately related, 
but the emphasis is different.

The exposition here is for an infinite system periodic in three dimensions.
This method has been implemented for two-dimensional systems,\cite{hjortstam0}
but that will not be discussed here.

\subsection*{Notation}
Papers on electronic structure methods unavoidably carry a high overhead in 
functional symbols and indices.
It is simplest to define here,
without motivation,
the special symbols and functions that will be used in this paper, for future reference.
These special functions 
(although not necessarily the symbols used here)
have been used extensively in LMTO documentation
and are largely due to Andersen.\cite{andersen}
\subsubsection*{Spherical harmonics:}
\begin{eqnarray}
      \Y_{\ell m}(\hat{\bvc r})
&
      \equiv
&
      \text{i}^{\ell} Y_{\ell m}(\hat{\bvc r})
\\
      C_{\ell m}(\hat{\bvc r})
&
      \equiv
&
      \sqrt\frac{4\pi}{2\ell\splus1}\ Y_{\ell m}(\hat{\bvc r})
\\
      \C_{\ell m}(\hat{\bvc r})
&
      \equiv
&
      \text{i}^{\ell} C_{\ell m}(\hat{\bvc r})
\end{eqnarray}
where $Y$ is a spherical harmonic.\cite{jackson0}
\subsubsection*{Bessel functions:}
\begin{eqnarray}
      \K_{\ell}(\kappa,r)
&
      \equiv
&
      - \kappa^{\ell\splus 1}
\begin{cases}
 n_{\ell}(\kappa r) -  \text{i} j_{\ell}(\kappa r) & \kappa^2 < 0 \\
 n_{\ell}(\kappa r)                        & \kappa^2 > 0 \\
\end{cases}
\\
      \K_{L}(\kappa,\vec r)
&
      \equiv
&
      \K_{\ell}(\kappa,r) \Y_{L}(\hat{\vec r})
\\
      \J_{\ell}(\kappa,r)
&
      \equiv
&
      j_{\ell}(\kappa r) / \kappa^{\ell}
\\
      \J_{L}(\kappa,\vec r)
&
      \equiv
&
      \J_{\ell}(\kappa,r) \Y_{L}(\hat{\vec r})
\end{eqnarray}
where $L$ denotes $\ell m$ and
$n_{\ell}$ and $j_{\ell}$ are spherical Neumann and Bessel functions,
respectively.

\subsubsection*{Geometry:}
For computational purposes, 
the crystal is divided into
non-overlapping spheres surrounding atomic sites
({\it muffin-tin spheres})
where the charge density and potential vary rapidly
and the {\it interstitial} region between the spheres,
where the charge density and potential vary slowly.
This is the {\it muffin-tin} geometry 
used as an idealized potential and charge density
in early electronic structure methods (KKR and APW).
Here, the division is a computational one,
and does not restrict the final shape of the charge density or potential.
In the muffin-tin spheres, 
the basis functions, electron density, and potential
are expanded in spherical waves;
in the interstitial region, 
the basis functions, electron density, and potential
are expanded in Fourier series.

There are many relevant considerations in choosing muffin-tin radii.
Assuming all expansions are taken to convergence,
the density and potential depend on the muffin-tin radii only through
the dependence of basis functions on the radii.
As discussed below, basis functions have a different functional form
inside the muffin-tin spheres, and the choice of muffin-tin radius affects this
crossover.
Hence, assuming the Hamiltonian is the same inside and outside the spheres
(the treatment of relativity may affect this as discussed below),
the muffin-tin radii are variational parameters 
and the optimum choice minimizes the total energy.
If the basis is large enough however (suitably complete within and without the spheres),
the energy is insensitive to the choice of radii.
A reasonable choice results from choosing radii that are 
within both the minimum in charge density and the maximum in potential along a line between nearest
neighbors.
Relativistic effects are usually taken into account only in the muffin-tin spheres,
in which case the Hamiltonian depends on the radii; 
hence when relativistic effects are important, the radii are not variational parameters.

In what follows, lattice positions are vectors $\vec{R} = R\vec{n}$, 
integer multiples of a basis $R$.
Atomic positions in the unit cell are denoted by $\vec{\tau}$.
A set of atomic positions invariant under the point group of the lattice 
are said
to be of the same symmetry type, $t$.
Similarly, in the reciprocal lattice, vectors are $\vec{g} = G\vec{n}$ for the 
reciprocal basis $G = 2\pi R^{-T}$.
Brillouin zone (or reciprocal unit cell) vectors are denoted by $\vec{k}$.

\subsubsection*{Symmetric functions:}

Within the muffin-tin region, functions invariant are expressed in harmonic series.
If $f(\vec{r})$ is such a function, at site $\tau$
\begin{subeqnarray}
      f(\vec r)\Big|_{r_{\tau}<s_{\tau}} 
&=&
      \sum_{h} f_{ht}(r_{\tau}) D_{ht}(\D_{\tau} \hat{\vec r}_{\tau})
\label{symfdef0}
\\
      D_{ht}(\hat{\vec r})
&=&
      \sum_{m} \alpha_{ht}(m) \C_{\ell_h m}(\hat{\vec r})
\label{dhtdef0}
\end{subeqnarray}
In Equation (\ref{symfdef0}), $\D_{\tau}$ is a transformation to a coordinate system
local to site $\tau$; 
the local coordinates of sites of the same type are related by an element of the
crystal point group that takes one site into another.
Expressed in this way, the functional form of $D_{ht}$ (Equation (\ref{dhtdef0})) 
depends only on symmetry type. 

In the interstitial region, symmetric functions are expressed in Fourier series:
\begin{subeqnarray}
      f(\vec r)\Big|_{r\in\I} 
&=&
      \sum_{\S} f(\S) D_{\S}(\vec r)
\label{symfdef1}
\\
      D_{\S}(\vec r)
&=&
      \sum_{g\in\S} e^{i\vec{g}\cdot\vec{r}}
\label{dsdef0}
\end{subeqnarray}
The sum in Equation (\ref{symfdef1}) is over symmetry stars $\S$ of the reciprocal lattice.

\section{Basis Set}
\label{sec.bases}

\subsection{Interstitial}

In the interstitial region 
(symbolically $\I$)
between the muffin-tin spheres, 
bases are Bloch sums of spherical Hankel or Neumann functions:
\begin{eqnarray}
\label{intbasis0}
      \psi_{i}(\bvc k,\bvc r) \Big|_{\bvc r\in\I}
&
      = 
&
      \sum_{R} e^{i\bvc k\cdot\bvc R}
      \K_{\ell_i}(\kappa_i,|\bvc r\sminus\boldtau_i\sminus\bvc R|)
      \Y_{\ell_{i}m_{i}}
            \bigl(\D_{\tau_{i}}(\bvc r\sminus\boldtau_i\sminus\bvc R)\bigr)
\end{eqnarray}
The rotation $\D_{\tau}$ in (\ref{intbasis0}) takes the argument into a
coordinate system local to each site $\tau$. 
The purpose of this will be made evident later.
The function on the right hand side of Equation (\ref{intbasis0})
is sometimes called the envelope function.

Notice the parameters, specifying a basis function, inherent in
this definition.
They are the site $\tau$ in the unit cell on which the spherical wave is
based, 
the angular momentum parameters $\ell$ and $m$ of the spherical wave
with respect to its parent cell,
and the kinetic energy $\kappa^2$ of the basis in the interstitial region.
The angular momentum parameters specifying the basis set 
are chosen to represent the atomic states from which crystal eigenstates
are derived.
In the LMTO-ASA, it is usual to include 
$\ell$ bases one higher than the highest relevant band.
In the method described here, this is rarely necessary, 
possibly because of the multiplicity of bases 
with the same angular momentum parameters.
It is usual to use ``multiple $\kappa$'' basis sets, 
having all parameters except the tail parameter the same.

There appears to be no simple algorithm for choosing 
a good set of interstitial kinetic energy parameters.
Schemes such as bracketing the relevant energy spectrum 
have been proposed.\cite{savrasov}
The optimum set would minimize the total energy.
This can be done but is time consuming even for relatively simple systems.
It seems, however that parameter sets obtained in this way for 
simple systems in representative configurations can give good results 
when used for related systems over a broad pressure range.
Thus good sets are arrived at through some experimentation.
The choice can be important as it's possible to pick a set of parameters
that will give very bad results, 
and the parameter set used in any new calculation should be always checked 
for stability.

\subsection{Muffin Tins}

In the muffin-tin spheres, 
bases are linear combinations of spherical waves 
matching continuously and differentiably to the envelope function
at the muffin-tin sphere.
The envelope function $\K$ 
may be expanded in a series of spherical Bessel functions 
about any site except it's center.
A basis function 
on a muffin-tin sphere in the unit cell at $\vec{R} = 0$ 
is therefore
\begin{eqnarray}
\label{mtbas1}
   \psi_{i}(\vec k,\vec r)
   \Big|_{r_{\tau} = s_{\tau}}
   = 
   \sum_{R} e^{i\vec k\cdot\vec R}
   \sum_{L}
   \Y_{L}(\D_{\tau} \hat{\vec r}_{\tau})
   \Bigl(
&&
         \K_{\ell}(\kappa_{i},s_{\tau})
         \delta(R,0) \delta(\tau,\tau_{i}) \delta(L,L_{i})
\nonumber \\
         +\ 
&&
         \J_{L}(\kappa,s_{\tau})
         B_{L,L_{i}}(\kappa_{i}, \vec\tau\sminus\vec\tau\Pr\sminus\vec R)
   \Bigr)
\nonumber \\
=
   \sum_{L}
   \Y_{L}(\D_{\tau} \hat{\vec r}_{\tau})
   \Bigl(
&&
         \K_{\ell}(\kappa_{i},s_{\tau})
         \delta(\tau,\tau_{i}) \delta(L,L_{i})
\nonumber \\
         +\ 
&&
         \J_{L}(\kappa,s_{\tau})
         B_{L,L_{i}}(\kappa_{i}, \vec\tau\sminus\vec\tau\Pr,\vec k)
   \Bigr)
\nonumber \\
&&
\end{eqnarray}
where 
$\vec{r}_{\tau} \equiv \vec{r} - \vec{\tau}$
and
$B$ is   equivalent to the KKR structure constant. \cite{kkrref}
The unitary transformation applied to $B$ rotates components into site-local
coordinates from the left and right.

Equation (\ref{mtbas1}) is compactly expressed by defining a two-component row vector
${\rm K}$ so that
\begin{eqnarray}
      {\rm K}_{\ell}(\kappa,r)
&
=
&
      \left(\K_{\ell}(\kappa,r), \J_{\ell}(\kappa,r)\right)
\end{eqnarray}
and a two component column vector ${\rm S}$ so that
\begin{eqnarray}
      {\rm S}_{L,L\Pr}(\kappa,\vec\tau\sminus\vec\tau\Pr,\vec k) 
&
=
&
      \left(
            \begin{matrix}
            \delta(\tau,\tau\Pr) \delta(L,L\Pr) \\
            B_{L,L\Pr}(\kappa,\vec\tau\sminus\vec\tau\Pr,\vec k) \\
            \end{matrix}
      \right)
\period
\label{defineS}
\end{eqnarray}
Then the value of a basis function on a muffin-tin boundary is 
expressed simply as
\begin{eqnarray}
   \psi_{i}(\vec k,\vec r)
   \Big|_{r_{\tau} = s_{\tau}}
& = &
   \sum_{L}
   \Y_{L}(\D_{\tau} \hat{\vec r}_{\tau})
   {\rm K}_{\ell}(\kappa_{i},s_{\tau})
   {\rm S}_{L,L_{i}}(\kappa_{i}, \vec\tau\sminus\vec\tau\Pr,\vec k)
\label{atbound0}
\end{eqnarray}

The radial part a basis function inside a muffin-tin sphere 
is a linear combination of atomic like functions
$\phi$ 
and their energy derivatives
$\dot\phi$
\cite{andersen,skriver}
matching continuously and differentiably to the radial function ${\rm K}$
in Equation (\ref{atbound0}).
Collecting $\phi$ and $\dot\phi$ in a row vector 
\begin{equation*}
      \RU(e,r) \equiv \left(\phi(e,r), \dot\phi(e,r) \right)\ \ ,
\end{equation*}
a simple case of
this matching condition may be expressed as
${\rm U}(e,s) \Omega(e,\kappa) = {\rm K}(\kappa,s)$
and
${\rm U}\Pr(e,s) \Omega(e,\kappa) = {\rm K}\Pr(\kappa,s)$,
where $\Omega$ is a matrix of order 2.

The use of these radial functions in the method described here
is different than that used by most other methods, however.
For the broadest utility, 
a basis set must be flexible enough to describe 
energy levels derived from atomic states having
different principle quantum numbers but 
the same angular momentum quantum number.
For example, describing the properties of elemental actinides at any pressure 
requires a basis with both $6p$ and $7p$ character.
Similarly, an adequate calculation of the structural properties of
transition metal oxides requires both semi-core and valence $s$ and $p$ states
on the transition metal ions.
The description of the evolution of core states from localized to itinerant
under pressure also requires multiple principle quantum numbers per $\ell$ 
value.
It is usual in LMTO-based methods
to perform calculations 
for the eigenstates and eigenvalues of ``semi-core'' and valence states
separately, 
using a different basis set, 
with a single set of energy parameters $\lbrace e_{\ell}\rbrace$,
for each ``energy panel''.
This approach fails when energy panels overlap, and
has the disadvantage that the set of eigenvectors is not
an orthogonal set.
The problem of ``ghost bands'' also arises.\cite{skriver} 

In the method described here,
bases corresponding to multiple principle quantum numbers 
are contained within a 
single, {\it fully hybridizing} basis set.
This is accomplished simply by using functions $\phi$ and $\dot\phi$ 
calculated with energies $\lbrace e_{n\ell}\rbrace$ 
corresponding to different principal quantum numbers $n$
to describe the radial dependence of a basis in the muffin-tin spheres.
The Hamiltonian matrix for an actinide,
for example,
will 
have elements 
$\big\langle \psi_{6p} \big | H \big | \psi_{7p} \big\rangle$
and the overlap matrix elements
$\big\langle \psi_{6p} \big | \psi_{7p} \big\rangle$,
We may formally express 
the radial part of basis $i$ in a muffin-tin sphere by 
the function
$f(r) = \sum_{n} a_{i}(n\ell) {\rm U}(e_{n\ell},s) 
      \Omega(e_{n\ell},\kappa_{i})$
but in practice 
it is sufficient to restrict the coefficients by 
$a_{i}(n\ell) = \delta(n,n_{i})$
so that the basis set  (although not eigenvectors)
will have pure principal quantum number ``parentage''.
This method of expanding the energy range of a basis set 
has been used (and reported) extensively.
Representative calculations in which this method was essential
are described in Reference \cite{willsused}.

Thus another parameter specifying a basis function 
is the set of energy parameters $\lbrace e_{t\ell} \rbrace$
that will be used to calculate the radial basis functions
$\phi_{t\ell}$ and $\dot\phi_{t\ell}$ 
used to express the basis function
in muffin-tin spheres of each symmetry type.
A basis function in a muffin-tin sphere is therefore
\begin{eqnarray}
      \psi_{i}(\vec k,\vec r) \Big|_{r_{\tau} < s_{t}}
& = &
      \sum_{L}^{\ell \leq \ell_{m}} 
               {\rm U}_{tL}(e_{i},\D_{\tau}\vec{r}_{\tau})
               \Omega_{t\ell}(e_{i},\kappa_{i})
               {\rm S}_{L,L_{i}}(\kappa_{i}, \vec\tau\sminus\vec\tau\Pr,\vec k)
\label{mtbasis0}
\end{eqnarray}
where 
$e_{i}$ means
``use the energy parameter $e_{n\ell}$ corresponding 
to the principal quantum number specified for basis i'' 
and 
\begin{equation}
      \RU_{tL}(e,\vec{r})
\equiv
      \Y_{L}(\hat{\vec{r}})
      \RU_{t\ell}(e,r)
\period
\end{equation}
The necessary cutoff in angular momentum has now been made explicit.
The $2\times 2$ matrix $\Omega$ matches ${\rm U}$ to ${\rm K}$ 
continuously and differentiably
at the muffin-tin radius.
Specifically, $\Omega$ is specified by
\begin{eqnarray}
      \left(\begin{matrix}
            \phi_{t\ell}(e,s_{t}) & \dot\phi_{t\ell}(e,s_{t}) \\
            \phi\Pr_{t\ell}(e,s_{t}) & \dot\phi\Pr_{t\ell}(e,s_{t})
            \end{matrix}\right)
      \Omega_{t\ell}(e,\kappa)
&=&
      \left(\begin{matrix}
            \K_{\ell}(\kappa,s_{t}) & \J_{\ell}(\kappa,s_{t}) \\
            \K\Pr_{\ell}(\kappa,s_{t}) & \J\Pr_{\ell}(\kappa,s_{t})
            \end{matrix}\right)
\end{eqnarray}

In principle, and as programmed,
each $(\tau \ell \kappa)$ basis can use
its own unique energy set.
It is more usual to use
a common energy set for a set of basis states 
giving rise to bands of similar energy 
within the scope of a particular calculation.
The configuration of the basis shown in Table~\ref{typbasis0}
for example uses a set of energies for ``semi-core'' $6s$ and $6p$ bases, 
and another set of energies to represent ``valence'' bases.
The calculation of energies in an energy parameter set
is discussed below.

A parameter introduced in (\ref{mtbasis0}) is the angular momentum
cut-off $\ell_{m}$.
In most cases, a converged total energy is achieved with values 
$\ell_{m} \sim 6 - 8$.
Note that since a basis set generally contains functions based on
spherical waves with $\ell \leq 3$, the KKR structure constant in
(\ref{defineS}) is rectangular.

\begin{table}
\caption{Parameters for typical basis set for an elemental actinide:
parent angular momentum parameter ($\ell$),
energy set for radial expansions ($e$-set),
and the index of the kinetic energy in the interstitial  region($\kappa$-index).
A typical set of $\kappa^2$ values, corresponding to the kinetic energy
indices, is given at the bottom of the table.
}
\renewcommand{\arraystretch}{1.1}
\setlength\tabcolsep{5pt}
\begin{tabular}{llll|llllllll}
\hline\noalign{\smallskip}
$n$ & $\ell$ & $e$-set & $\kappa$-index & 
    $n$ & $\ell$ & $e$-set & $\kappa$-index & 
    $n$ & $\ell$ & $e$-set & $\kappa$-index \\
\hline
6 & $s$ & 1 & 1 & 7 & $s$ & 2 & 3 & 6 & $d$ & 2 & 3 \\
6 & $s$ & 1 & 2 & 7 & $s$ & 2 & 4 & 6 & $d$ & 2 & 5 \\
6 & $p$ & 1 & 1 & 7 & $s$ & 2 & 5 & 5 & $s$ & 2 & 3 \\
6 & $p$ & 1 & 2 & 7 & $p$ & 2 & 3 & 5 & $s$ & 2 & 5 \\
  &     &   &   & 7 & $p$ & 2 & 4 \\
  &     &   &   & 7 & $p$ & 2 & 5 \\
\hline
$\kappa^2$: & \multicolumn{3}{l}{1:\quad $-$1.96582916} &
                  & \multicolumn{3}{l}{3:\quad $-$3.44402161} \\
            & \multicolumn{3}{l}{2:\quad $-$.193652690} &
                  & \multicolumn{3}{l}{4:\quad $-$1.56582916} \\
            & \multicolumn{3}{l}{\hskip 2ex} &
                  & \multicolumn{3}{l}{5:\quad  .331719550} \\
\hline
\end{tabular}
\label{typbasis0}
\end{table}

\section{Matrix Elements}

\subsection{Muffin-Tin Matrix Elements}

The potential in a muffin-tin at $\vec\tau$ 
has an expansion in linear combinations of spherical harmonics 
invariant under 
that part of the point group leaving $\vec\tau$ invariant:
\begin{subeqnarray}
\label{defpotmt0}
      V(\bvc r)\Big|_{r_{\tau}<s_{t}}
&=& 
      \sum_{h} v_{ht}(r_{\tau}) 
            D_{ht}\bigl( \D_{\tau} \hat{\vec r}_{\tau} \bigr)
\\
         D_{ht}(\hat{\vec r}) 
&=&
         \sum_{m} \alpha_{ht}(m) \C_{\ell_{h}m}(\hat{\vec r})
\label{dhtdef1}
\period
\end{subeqnarray}
The utility of referring bases and potentials in muffin-tin spheres 
to site-local coordinates is apparent in (\ref{defpotmt0}).
If the site local coordinates of sites are constructed so that
$\D_{\tau\Pr} = \D_{\tau} {\cal Q}^{-1}$ for some ${\cal Q}$ 
such that ${\cal Q} \boldtau = \boldtau\Pr$, then
the harmonic functions $D_{ht}$ depend only on the symmetry type,
rather than on each site.
The normalization for the spherical harmonic in (\ref{defpotmt0})
($\C = \sqrt{4\pi/(2\ell\splus1)} \Y$)
is chosen so that $v_{ht}(r)$ is the potential when $\ell_{h} = 0$.

Combining (\ref{mtbasis0}) and (\ref{defpotmt0}), the potential matrix
is
\begin{eqnarray}
      \bra{\psi_{i}} 
      V 
      \ket{\psi_{j}} \Big|_{mt}
      &=&
      \sum_{\tau}
      \sum_{L}
         {\rm S}^{\dagger}_{L,L_{i}} 
               (\kappa_{i},\vec\tau\sminus\vec\tau_{i},\vec k)
\\
      &\times&
      \Bigl(
      \sum_{h}
      \sum_{L\Pr}
             \Omega^{T}_{t \ell}(e_{i},\kappa_{i})
             \bra{U^{T}_{t\ell}(e_{i})} v_{ht} \ket{U_{t\ell}(e_j)}
             \Omega_{t \ell}(e_{j},\kappa_{j})
\nonumber
\\
&& \hskip 3.3 cm
             \bra{L} D_{ht} \ket{L\Pr}\,
             {\rm S}_{L,L_{j}} 
                   (\kappa_{j},\vec\tau\sminus\vec\tau_{j},\vec k)
      \Bigr)
\period
\nonumber
\end{eqnarray}
The matrix element of the $D_{ht}$ is a sum over Gaunt coefficients:
\begin{eqnarray*}
\bra{L} D_{ht} \ket{L\Pr}
&=&
\sum_{m_{h}} \alpha_{ht}(m_{h}) 
             \gauntcof{\ell\Pr}{m\Pr}{\ell}{m}{\ell_{h}}{m_{h}}
\\
\gauntcof{\ell\Pr}{m\Pr}{\ell}{m}{\ell_{h}}{m_{h}}
&=&
\int \Y_{\ell\Pr m\Pr} \Conjg{\Y}_{\ell m} \C_{\ell_{h} m_{h}}
\\
\end{eqnarray*}

In electronic structure methods using muffin-tin orbitals,
the muffin-tin energy parameters $\lbrace e_{\ell}$
are usually taken from ``$\ell$-projected average energies''.
With multiple energy sets, this is a reasonable choice provided that 
the basis set, which uses separate sets, gives rise to bands well separated 
in energy.
The $\ell$-projected charge, integrated over a muffin-tin sphere, is a 
sum over cross terms between energy sets 
\begin{equation*}
      Q_{\ell} = \sum_{ij} Q_{\ell}(e_{i},e_{j})
\end{equation*}
and must be made diagonal in some approximation for the 
resulting energy- and $\ell$-projected energies and charges to be
representative.

Another criterion, particularly useful for states using different sets
not well separated in energy or for states not having significant occupation
is to maximize the completeness of the basis.
To accomplish this, the energy parameter for the low energy state
$e_{\ell}(1)$ can be set to a set of projected energy averages, and the
energy parameters for the same $\ell$ in higher energy sets may be 
chosen so that the radial function has one more node and the same
logarithmic derivative at the muffin-tin radius, hence
\begin{equation}
\int_0^s r^2 dr\ \phi_{\ell}(e_{1},r) \phi_{\ell}(e_{i},r)
=
0
\comma\ \ i > 1
\period
\label{en_orth_cond}
\end{equation}
Although this usually generates energy parameters out of the range
of occupied states (since the logarithmic derivative of semi-core
states is usually large in magnitude and negative), 
this choice seems to give a total energy close to the minimum with 
respect to this parameter.
This is an example of the difference mentioned in the introduction 
in emphasis between an accurate ``basis-set'' method and a 
method motivated by a
physical model.

The convergence of the harmonic expansion of the potential in a muffin-tin
sphere (\ref{defpotmt0}) depends, of course, on the basis, atomic constituents, 
and geometry.
Using harmonics through $\ell_{h_{max}} = 6$ is usually sufficient, and it
has never been necessary to go beyond $\ell_{h_{max}} = 8$.

\subsection{Interstitial Matrix Elements}
\label{inmats}

\subsubsection{Overlap and Kinetic Energy:}

The interstitial overlap matrix can be easily obtained from an
integral over the interstitial surface 
(the only non-zero contributions, 
in a crystal periodic in three dimensions, 
come from the surfaces of the muffin-tin spheres)
and the kinetic energy is proportional to the overlap:
\begin{eqnarray}
      \int_{\I} \psi_{i}^{\dagger}(\vec r) \psi_{j}(\vec r)
      &=&
      -
      (\kappa_j^2 - \kappa_i^2)^{-1}
      \int_{\I} \bigl(          \psi_{i}^{\dagger} \nabla^2 \psi_{j}
                    - (\nabla^2 \psi_{i}^{\dagger})         \psi_{j} \bigr)
\nonumber \\
      &=&
      (\kappa_j^2 - \kappa_i^2)^{-1}
      \sum_{\tau}
      s_{t}^2
      \int d\Omega_{\tau}
      W(\psi_{i}^{\dagger},\psi_{j})
\label{intovlp0}
\end{eqnarray}
where $W(f,g) = f g\Pr - f\Pr g$.
Basis functions on muffin-tin spheres are given in (\ref{atbound0}), hence
\begin{eqnarray}
\label{inov1}
      \braket{\psi_{i}}{\psi_{j}}\Big|_{\I} 
=
      \sum_{\tau} s_{t}^{2} \sum_{L} 
&&
      {\rm S}^{\dagger}_{L,L_{i}}(\kappa_{i}, \vec\tau\sminus\vec\tau_{i},\vec k)
\nonumber \\
      \times 
&&
      \frac{ W\big({\rm K}^{T}_{\ell}(\kappa_{i},s_{t}),{\rm K}_{\ell}(\kappa_{j},s_{t}) \bigr)
      }{
      \kappa_{j}^{2}-\kappa_{i}^{2} }
      {\rm S}_{L,L_{j}}(\kappa_{j}, \vec\tau\sminus\vec\tau_{j},\vec k)
\end{eqnarray}
In the limit $\kappa_j^2 \to \kappa_i^2$, 
the evaluation of (\ref{intovlp0}) requires
the derivative with respect to $\kappa^2$ of the structure constant.

\subsubsection{Potential Matrix Elements:}
The greatest difference between LMTO-based full-potential methods
is in the way the matrix elements of the potential are calculated over
the interstitial region.
The method being described here uses a Fourier representation of basis
functions and the interstitial potential to calculate these matrix
elements. 
Other approaches for computing these elements are described in 
the literature. \cite{springborg,methfessel}

A Fourier transform of the basis functions described in Section \ref{sec.bases} 
would be too poorly convergent for practical use.
However, the evaluation of the interstitial potential matrix
requires only a correct treatment of basis functions and potential in 
the interstitial region.
This degree of freedom can be  used to design ``pseudo basis-set'', equal 
to the 
true basis in the interstitial region although not in the muffin-tin spheres, 
and have a Fourier transform which converges rapidly enough for practical use.
We define this pseudo basis set  by
\begin{subeqnarray}
      \tilde\psi_{i}(\vec k,\vec r) \Big|_{\vec r\in\I}
&=&
      \sum_{R} e^{i\vec k\cdot\vec R}
      \tilde{\cal K}_{\ell_i}(\kappa_i,|\vec r\sminus\boldtau_i\sminus\vec R|)
      i^{\ell} Y_{\ell_{i}m_{i}}(\vec r\sminus\boldtau_i\sminus\vec R)
\\
      \tilde\K_{\ell}(\kappa,r)
&\equiv&
      \K_{\ell}(\kappa,r), \ \ r > s,\ \ s \leq s_{\tau}
\end{subeqnarray}

Since rapid Fourier convergence is the criterion for constructing the
pseudo-basis, it is useful to consider the Fourier integral of a 
Bloch function with wave-number $\bvc k$:
\begin{equation}
      \tilde\psi(\vec g)
=
      -
      \frac{1 }{ {\rm V}_{c} (|\bvc k\splus\vec g|^2 - \kappa^2)}
      \int_{{\rm V}_{c}} d^3r\, 
            e^{-i(\vec k\splus\vec g)\cdot\vec r}
            \bigl(\nabla^2 \splus \kappa^2)
            \tilde\psi(\vec r)
\label{fourier0}
\end{equation}
where V$_c$ is the unit cell volume.
Equation (\ref{fourier0}) is obtained by casting $\nabla^2 \splus \kappa^2$
on the plane wave then doing two partial integrations; surface terms vanish
due to periodicity.
From (\ref{fourier0}) it is evident that the Fourier integral of a pseudo-basis
satisfying the first criterion (equal to the true basis in the interstitial
region)
may be obtained from integral over muffin-tin spheres.
If in addition, the pseudo-basis is different from a Hankel function only
in it's parent sphere, the Fourier integral is a finite integral over a
single muffin-tin sphere.
The problem then is to find a function $\tilde\psi$ 
such that $(\nabla^2\splus\kappa^2) \tilde\psi$ 
has a rapidly convergent Fourier integral, vanishes outside a radius less
than or equal to the parent muffin-tin radius for the basis, and has
a value and slope equal to $\K$ at this radius.

A good choice for such a function is obtained by solving
\begin{equation}
      \left(\nabla^2 + \kappa^2\right) 
            \tilde\K_{\ell}(\kappa,r)
            \Y_{L}(\hat{\vec r})
      =
      - c_{\ell}
      \Bigl( \frac{r}{ s} \Bigr)^{\kern-2pt\ell}
      \Bigl[ 1 - \Bigl(\frac{r}{s}\Bigr)^{\kern-2pt 2} \Bigr]^{n}
      \Y_{L}(\hat{\vec r})
      \Theta(s-r)
\label{pseudo0}
\end{equation}
for a radius $s < s_{t_{i}}$, and with
with $c_{\ell}$ chosen to match on to $\K$ at $s$.
This is easily done analytically.
The resulting Fourier transform is
\begin{equation}
      \tilde\psi_{i}(\vec k\splus\vec g)
=
      \frac{4 \pi }{ {\rm V}_{c} }
      \frac{
            Y_{L_{i}}(\vec k\splus\vec g) 
            e^{-\text{i}(\vec k\splus\vec g)\cdot\vec\tau_{i}}
      }{
            \left(|\vec k\splus\vec g|^2 - \kappa_{i}^2\right)
      }
      |\vec k\splus\vec g|^{\ell_{i}}
      \frac{
            \J_{N}(|\vec k\splus\vec g|, s)
      }{
            \J_{N}(\kappa_{i}, s)
      }
\label{pseudo1}
\end{equation}
where $N = \ell_i\splus n_{i}\splus1$.
The subscript $i$ has been purposely left off $N$ and $s$ (see below).

These coefficients converge like $1/g^{n+4}$, 
provided $\J_{N}(|\vec k\splus\vec g|,s)$ achieves it's large argument behavior,
and $n$ can be chosen to optimize convergence.
Weinert \cite{weinert0} used an analogous construction as tool  to 
solve Poisson's equation.
He proposed a criterion for the convergence of the Fourier serie 
(\ref{pseudo1}) which amounts to choosing the
exponent $n$ in Equation (\ref{pseudo1}) so that $|\vec k\splus\vec g_{max}|s$
would be greater than the position of the first node of 
$\J_{\ell\splus n\splus1}$.
We find this criterion to be useful provided anisotropy in reciprocal space
is accounted for.
This is accomplished by using the minimum reciprocal lattice vector
on the surface of maximal reciprocal lattice vectors, rather than simply
using $g_{max}$.

Notice that this criterion is a criterion for $N = \ell\splus n\splus1$.
The basis Fourier components are simplified, and the amount of information
stored reduced, by simply using a single argument for all bases; 
{\it i.e.} all bases use the same value of $N$.
It is also possible to use a single radius $s$, less than or equal to
the smallest muffin-tin radius, since the only requirement is on the
pseudo bases in the interstitial region.
In practice, a few radii are desirable if large and small atoms are
present in the same calculation, since small radii give less convergent
Fourier coefficients.
In any event, no more than a few radii are necessary to handle systems with
many atoms.
Notice also that local coordinates have been left out of (\ref{pseudo1}).
The resulting potential matrix may be easily rotated to local coordinates
at the end of the calculation.

As expressed in (\ref{pseudo1}), the Fourier components are products of
phases $e^{-\text{i}(\vec k\splus\vec g)\cdot\vec{\tau}}$, which scale like the number
of atoms squared (the size of the reciprocal lattice grid grows linearly
with the number of atoms), and a function of lattice vectors and a few
parameters, which scales linearly with the number of atoms.
The phase factors are simple to calculate by accumulation and need not be
stored.

The potential in the interstitial region is similarly obtained from a 
``pseudo-potential'' $ \tilde V$ that  equals the true
potential in the interstitial region and has rapidly converging Fourier 
coefficients:
\begin{subeqnarray}
      V(\vec r) \Big|_{\I}
      &=& 
      \tilde V(\vec r) \Big|_{\I}
\\
      \tilde V(\vec r) 
      &=&
      \sum_{{\cal S}} \tilde V({\cal S}) D_{{\cal S}}(\vec r)
\label{foupot0}
\\
      D_{{\cal S}}
      &=&
      \sum_{\vec g\in{\cal S}} e^{\text{i}\vec g\cdot\vec r}
\end{subeqnarray}
The sum in Equation (\ref{foupot0}) is over stars ${\cal S}$ of the reciprocal lattice.

Integrals over the interstitial region are performed by convoluting the
potential with an interstitial region step function and integrating over the
unit cell:
\begin{eqnarray*}
      \bra{\psi_{i}} V \ket{\psi_{j}}_{\I}
&=&
      \bra{\tilde\psi_{i}} \tilde V \ket{\tilde\psi_{j}}_{\I}
      = 
      \bra{\tilde\psi_{i}} \theta_{\I} \tilde V \ket{\tilde\psi_{j}}_{c}
\period
\end{eqnarray*}
The potential matrix element is calculated by convoluting the convoluted
potential with a basis, and performing a direct product between convoluted
and unconvoluted bases.
If basis functions are calculated $n^{3}$ reciprocal lattice vectors, the
interstitial potential will be calculated on $(2n)^{3}$ vectors.
The convolution is exact if it is carried out on a lattice containing
$(4n)^{3}$ vectors.
The size of the set of reciprocal lattice vectors necessary to converge 
the total energy using this
treatment of the interstitial region varies from between $\sim$ 150 -- 300
basis plane waves per atom, depending on the smoothness of the potential and
the convergence required.

Another
way of integrating over the interstitial region, 
more usual in site-centered methods, 
is to 
integrate Fourier series over the unit cell and subtract the muffin-tin
contributions with pseudo-bases and pseudo-potential expressed as an
expansion in spherical waves.
The convolution has an advantage in acting with a single representation,
and, given a finite representation for bases and potential, the convolution
may be done exactly.

Empty spheres are never used with this scheme.  
Bases, and the charge density and potential are calculated as 
accurately as necessary 
using the scheme described above
and a basis set expanded with tail parameters and energy sets has proven to
be flexible enough 
to accurately describe the contribution of the electronic states 
in the interstitial region.

\section{Charge Density}

When a solution to the wave equation at every physical energy
is available, 
the charge density may be obtained from a set of energy-dependent
coefficients. 
The spherically symmetric charge density in a muffin-tin sphere,
coupled with an $\ell-projected$ density of states, 
is an example.
In a variational calculation, as is being described here, 
all that is available is a (variational) solution to the wave equation 
at a set of discreet energies, and the charge density must be obtained
simply from the square of the eigenvectors, 
or equivalently from expectation values of occupation numbers.

Having calculated a set of eigenvalues and eigenvectors ${\cal A}$ 
of the generalized
eigenvalue problem, the charge density in the interstitial region is 
\begin{subeqnarray}
      \tilde n(\vec r) \Big|_{\I}
&=& 
      \sum_{\S} \tilde n(\S) D_{\S}(\vec r)
\\
      \tilde n(\S)
&=&
      \frac{1 }{ N_{\S}}
      \sum_{\vec g\in\S}
      \sum_{nk} w_{nk} 
            \frac{1 }{ {\rm V}_c} \int_{{\rm V}_c} d^{3}r\,
            e^{-\text{i}\vec g\cdot \vec r} 
            \big| \sum_{i} \tilde\psi_{i}(k,\vec r)
                                     {\cal A}_{i}(nk) \big|^2
\end{subeqnarray}
where $N_{\S}$ is the number of vectors in the reciprocal lattice star $\S$.
The square of the wave function is obtained by convoluting the 
Fourier components of $\psi$ with ${\cal A}$, Fourier transforming, and 
taking the modulus.

In the muffin-tin spheres the charge density is
\begin{subeqnarray}
      n(\vec r)\Big|_{r_{\tau}<s_t}
      &=& 
      \sum_{h} n_{ht}(r_{\tau}) 
              D_{ht}( \D_{\tau} r_{\tau} )
\\
      n_{ht}(r)
      &=&
      \sum_{e\ell} \sum_{e\Pr\ell\Pr}
          U_{t\ell\Pr}(e_{i\Pr},r) 
          M_{ht}(e\ell,e\Pr\ell\Pr)
          U^{T}_{t\ell}(e_{i},r) 
\\
          M_{ht}(e\ell,e\Pr\ell\Pr)
&=&
      \frac{2\ell_{h}\splus1 }{ 4\pi}
      \sum_{m_{h} m m\Pr} 
      \Conjg{\alpha}_{ht}(m_{h})
      \gauntcof{\ell}{m}{\ell\Pr}{m\Pr}{\ell_{h}}{m_{h}}
\\
& & \hskip 50pt \times
      \sum_{nk} w_{nk}
          {\cal V}_{\tau\ell m}(e)
          {\cal V}^{\dagger}_{\tau\ell\Pr m\Pr}(e\Pr)
\nonumber
\\
          {\cal V}_{\tau\ell m}(e)
&=&
          \sum_{i}
          \delta(e,e_{i})
          \Omega_{t\ell}(e,\kappa_{i}) 
          {\rm S}_{\ell m, \ell_{i} m_{i}}
                (\kappa_{i},\vec\tau\sminus\vec\tau_{i},\vec k)
          {\cal A}_{i}(n\vec k)
\end{subeqnarray}
The process of calculation is evident in the sequence of equations.

\section{Core States}

Core states, even spherically symmetric complete shells,
contribute non-muffin-tin components to the interstitial region and to muffin-tin
spheres
surrounding other sites.
Whether it is essential to include this contribution depends on the 
size of the contribution,
and any sizable contribution implies that there are states being treated as localized
which aren't localized within the scope of the calculation.
Nevertheless, confining states to the core is often useful, and
including the core contribution to the full potential is not difficult.
One possibility, the one used in this method, is to fit the part of the core
electron density to a linear combination of Hankel functions, and expand 
this density in the interstitial region as a Fourier serie 
and in the muffin-tin spheres in a harmonic series, 
in the same
way the basis functions are treated.

\section{Potential}

\subsection{Coulomb Potential}
The Coulomb potential is obtained by 
first calculating the Coulomb potential in the interstitial region, 
then, using the value of the interstitial potential on the muffin-tin sphere, 
calculating the potential in the spheres by a numerical 
Coulomb integral of the muffin-tin electron density for each harmonic.

The interstitial Coulomb potential is calculated in a way similar to that 
suggested by Weinert  \cite{weinert0}.
Express the electron density as
\begin{eqnarray}
      n(\vec r) 
      &=&
      \tilde n(\vec r) 
      + 
      \sum_{R\tau} \bigl( n({\vec r}) - \tilde n(\vec r) \bigr)
           \Theta(s_{t}-r_{\tau})
\label{potential0}
\end{eqnarray}
where $\tilde n$ is the squared modulus of the pseudo-eigenvectors,
which is equal to the true electron density in the interstitial region.
The first term on the right-hand side of (\ref{potential0}) has, 
by construction, a convergent Fourier series.  
The second term is confined to  muffin-tin spheres.
To calculate the Coulomb potential in the interstitial region, this term
may be replaced by any density also confined to the muffin-tin spheres and 
having the same multipole moments.
If a charge density satisfies these requirements and also has a convergent
Fourier series, the Coulomb potential in the interstitial region may be easily
calculated from the combined Fourier series.
Such a charge density can be constructed in a similar way to that detailed 
for the pseudo-bases.
Construct a pseudo charge-density satisfying 
\begin{subeqnarray}
      \tilde n^{(p)}(\vec r)
      &=&
      \sum_{R\tau} \sum_{h} 
      \tilde n^{(p)}(ht,r_{R\tau}) D_{ht}(\D_{\tau}\hat{\vec{r}_{R\tau}})
\\
      \tilde n^{(p)}_{ht}(r) 
      &=& 
      c_{ht} 
      \Bigl( \frac{r}{s_{t}} \Bigr)^{\kern-1pt\ell_{h}}
      \Bigl( 1 - \Bigl(\frac{r}{s_{t}}\Bigr)^{\kern-2pt 2}\Bigr)^{\kern-1pt n}
      \Theta(s_t\sminus r)
\\
      0
      &=&
      \int_{\tau} d^{3}r\,
      r_{\tau}^{\ell} \Conjg{D}_{ht}(\D_{\tau} \hat{\vec r}_{\tau})
      \bigl(
            \tilde{n}^{(p)}(\vec r) \sminus n(\vec r) 
           \splus \tilde{n}(\vec r)
      \bigr)
\period
\end{subeqnarray}
This charge density has Fourier components
\begin{eqnarray}
      \tilde n^{(p)}(\vec r)
&=&
      \sum_{\tau}
      \sum_{h}
      e^{-\text{i}\vec{g}\cdot\vec{\tau}}
      (-i)^{\ell_{h}} D_{ht}(\D_{\tau}\vec{g})
      \frac{4\pi}{{\rm V}_{c}}
      \frac{(Q_{ht}\lbrace n\rbrace \sminus Q_{ht}\lbrace \tilde n\rbrace) 
            }{s^{\ell_{h}\splus n\splus 1}} 
\nonumber\\
&& \hskip 10 pt\times\ 
      \frac{
       \bigl(2(\ell_{h}\splus n\splus 1)+1\bigr)\Dblfac
       }{
       (2\ell_{h}+1)\Dblfac
      }
      g^{\ell_{h}}
      \J_{\ell_{h}\splus n\splus 1}(g,s_t)
\end{eqnarray}
where the multipole moments $Q$ are defined by
\begin{equation}
      Q_{ht}\lbrace n\rbrace
      = \frac{2\ell_{h}\splus 1 }{ 4\pi}
        \int_{s_{t}>r_{\tau}} 
           r_{\tau}^{\ell_{h}} D_{ht}(\hat{\vec{r}_{\tau}}) n(\vec{r})
           \ d^{3}\kern-1pt r_{\tau} 
\end{equation}
The Fourier components $\tilde n^{(p)}(\vec r)$
converge like $1 / g^{n+2}$ provided $j_{\ell\splus n\splus1}$
attains it's asymptotic form.
The exponent $n$ is chosen using the same considerations as for the 
pseudo-basis set.

The Coulomb potential in the interstitial region is then given by
\begin{eqnarray}
      V_{c}(\vec r) \Big|_{\I} 
&=&
      \tilde{V}_{c}(\vec r) \Big|_{\I} 
\nonumber
\\
&=&
      \sum_{g\neq 0} \frac{4\pi e^2 \bigl(\tilde n(g) \splus n^{(p)}(g) \bigr)
                          }{ g^2}
                     e^{\text{i}\vec g\cdot\vec r}
\end{eqnarray}

From the Coulomb potential in the interstitial region follows the Coulomb
potential on the surface of the muffin-tin spheres.
The coulomb Potential inside the muffin-tin spheres is 
\begin{eqnarray}
      V^{(c)}(\bvc r) \Big|_{r_{\tau}<s_{t}}
&
      =
&
      \sum_{h} D_{ht}(\D_{\tau}\hat{\bvc r_{\tau}})
      \Bigl[
      e^2
      \int_{0}^{s_{t}} 
         \frac{r^{\ell_{h}}_{<} }{ r^{\ell_{h}\splus1}_{>}}
         \frac{4\pi r\Prpow{2} n_{h}(r) }{ 2\ell_{h}\splus 1}
         dr\Pr
\\
&
      +
&
      \Bigl(
         V^{(c)}_{h}(s) 
         - 
         \frac{e^2 }{ s^{\ell_{h}\splus 1}}
         \int_{0}^{s}     
         \frac{4\pi r\Prpow{\ell_{h}\splus 2} n_{h}(r\Pr) }{ 2\ell_{h}\splus 1}
         dr\Pr
      \Bigr)
      \Bigl(\frac{r }{ s}\Bigr)^{\ell_{h}}
      \Bigr]
\nonumber
\end{eqnarray}
where 
\begin{equation}
      V^{(c)}_{ht}(s_{t})
\equiv
      \frac{2\ell_{h}\splus1}{4\pi}
      \int_{r_{\tau}=s_{t}} d\hat{\vec r}
      \Conjg{D}_{ht}(\D_{\tau}\hat{\vec r})
      V^{(c)}(\vec r)
\end{equation}
is the harmonic component of the potential on a sphere boundary.

\subsection{Density Gradients}
Gradients of the electron density are needed for the evaluation of gradient 
corrected density functionals. 
These functionals depend on invariants 
(with respect to the point group)
constructed from density gradients ({\it e.g.} $|\vec\nabla n|^{2}$).
This reduces computation significantly in the muffin-tin spheres, for
if $f$ and $g$ are invariant functions ({\it i.e.} $f(\vec r) = \sum_{h} f_{h}(r) D_{h}(\hat{\vec{r}})$),
and $d = \vec{\nabla}f \cdot \vec{\nabla} g$, then $d(\vec{r}) = \sum_{h} d_{h}(r) D_{h}(\hat{\vec r})$ with
\begin{eqnarray}
      \frac{4\pi r^2 }{ 2\ell_h\splus 1} d_{h}(r) &=& \sum_{h,h\Pr} \sum_{k,k\Pr = \pm 1}
            f^{(k)}_{h}(r) g^{(k\Pr)}_{h\Pr}(r) I(kk\Pr;hh\Pr)
\end{eqnarray}
where the set of parameters $I$ is easily calculable from $3j$ and $6j$ coefficients and
integrals over the harmonic functions $D_h$, and 
\begin{eqnarray}
\label{radder1}
      f^{(k)}_h 
&=&
\frac{4\pi }{ 2\ell\splus1}
\begin{cases}
r f\Pr - \ell_{h} & k = 1 \\
r f\Pr + \ell_h + 1 & k = -1 \\
\end{cases}
\end{eqnarray}
and similarly for $g$.

Gradients of the interstitial charge density, represented as a Fourier series,
are poorly represented by differentiating the series term by term.
A stable representation of the density gradient that converges well is obtained by 
defining the derivative as the difference between adjacent grid points,
divided by twice the grid spacing as suggested by Lanczos.\cite{lancsos0}
This is equivalent to differentiating, term by term, the Lanczos-damped series
for the charge density.

\section{All-Electron Force Calculations}
\subsection{Symmetry}
The set of internal forces acting on the atomic sites of a crystal 
is a symmetric, discrete function of atomic coordinates and
has a spherical expansion on the crystal sites with the same coefficients as 
continuous symmetric functions  (\ref{symfdef0}) and (\ref{dhtdef0}).
Since forces are vectors, their representation has $\ell = 1$,
and if a site has no invariant harmonics with $\ell = 1$,
there is no force on that site.
So the force on an atomic site may be expressed as
\begin{eqnarray}
\label{force0}
      \vec{f}(\tau)
&=&
      \sum_{h:\ell_{h}=1} f_{ht} \sum_{m} \alpha_{m} \hat{\vec{\rm e}}_m\ {\cal U}_{\tau}
\end{eqnarray}
where the coefficients $\alpha$ are as in  (\ref{dhtdef0}), 
the $\hat{\vec{\rm e}}_{m}$ are spherical unit vectors, \cite{edmonds0} and 
${\cal U}_{\tau}$ is the transformation to local coordinates for spherical vectors.
A force calculation is, as much as possible, a calculation of the set $\lbrace f_{ht} \rbrace$;
The size of this set is often much smaller than three times the number of atoms.
The displacements of atoms allowed by symmetry also have the form of   (\ref{force0}):
\begin{eqnarray}
\label{displ0}
      \delta \vec{\tau}
&=&
      \sum_{h:\ell_{h}=1} \delta\tau_{ht} \sum_{m} \alpha_{m} \hat{\vec{\rm e}}_m\ {\cal U}_{\tau}
\end{eqnarray}

Minimizing the energy with respect to the atomic  positions is a 
process of finding the set 
$\lbrace \delta\tau_{ht} \rbrace$ that gives $f_{ht} = 0$.

\subsection{Force Calculations}

\begin{figure}[t]
\includegraphics[width=.7\textwidth]{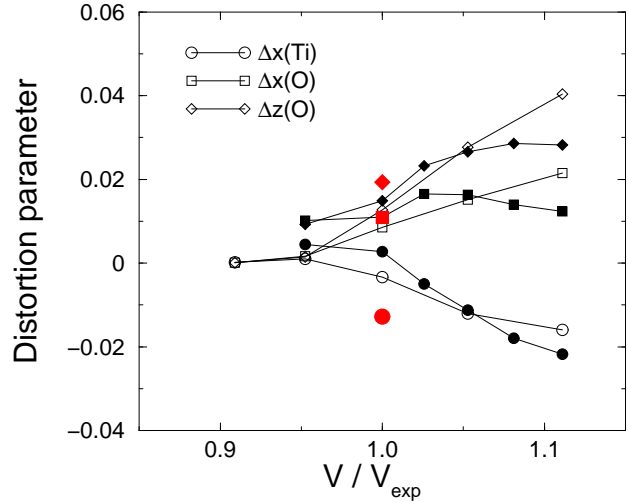}
\caption{The deviation of the internal coordinates of rhombohedral BaTiO$_{3}$ from ideal, 
calculated using all-electron force calculations as a function of volume 
with both LDA (open symbols) and GGA (filled symbols) exchange-correlation functions.  
The grey filled symbols are experimental points\protect{\cite{kwei}}.
The LDA equilibrium volume is .958 V$_{exp}$; the GGA volume is 1.037 V$_{exp}$.
The energy was also minimized with respect to the rhombohedral angle at each volume.}
\label{dist_parms}
\end{figure}

The calculation of forces in an all-electron method has been nicely described by 
Yu {\it et al.} \cite{Yu0} for the LAPW method. 
In addition to the terms discussed in that paper,
a force calculation using a site-centered basis has the additional,
and significant, complication that the bases depend on atomic position not
only through augmentation but also through parentage.

The contributions to the total force on a site in an all-electron calculation
follow directly from a derivative of the LDA total energy with respect 
to atomic positions.
The terms listed by Yu {\it et al.} are 1) a ``Helmann-Feynman'' term, 
$\partial E/\partial \vec\tau$, 
which accounts for the explicit dependence of the energy functional
on atomic positions,
2) an ``Incomplete Basis Set'' (IBS) term, which arises when derivatives of basis functions
aren't contained in the space covered by the basis set, 
3) a core-correction term, arising because core states 
are calculated using only the spherical average of the potential, and
4) a muffin-tin term, a surface term arising from the change in integration
boundaries when atoms are moved and the discontinuity of the second derivative of
basis functions across muffin-tin boundaries.
There are two other terms to consider.
The first arises when a calculation isn't fully self-consistent, and has the form
$- \int_{{\rm V}_{c}} (V_{\rm out} - V_{\rm in}) dn(\vec{r})/d\vec{\tau}$,
where $V_{\rm out}$ and $V_{\rm in}$ are output and input potentials.
The second term arises from the way in which Brillouin zone integrals are done.
Whether by quadrature or linear interpolation, the result is a set of weights
(occupations) multiplying quantities evaluated at discrete Brillouin zone 
points.  The terms listed above do not take into account the change 
of weights with atomic positions.

The evaluation of the IBS term in a method using site-centered bases 
is significantly more involved than in the LAPW method.
This term has the form
\begin{eqnarray}
\label{ibsform0}
\vec{F}_{\rm IBS}
=
-
\sum_{n\vec k} w_{n\vec k} \sum_{ij}
{\cal A}^{\ast}_{i,n\vec{k}}
&\Bigl(&
      \bra{\psi_{i}} H - e_{n\vec{k}} \ket{d\psi_{j}/d\vec\tau} 
\nonumber\\
      &+&
      \bra{d\psi_{i}/d\vec\tau} H - e_{n\vec{k}} \ket{\psi_{j}} 
\Bigr)
{\cal A}_{j,n\vec{k}}
\end{eqnarray}
where the $\cal A$ are eigenvectors.
Both LAPW and LMTO methods have a dependence on atomic positions through augmentation 
(the expansion of the basis set in atomic-like spherical waves)
in the muffin-tin spheres, 
and both methods have an implicit dependence of basis functions
on atomic positions through self-consistency, a term largely ignored 
and usually negligible.
A site-centered basis, however, depends on atomic positions also through 
it's parent site (the site it's centered on).
The contribution from augmentation is fairly easily accounted for at the 
density stage
of a calculation, after integrals over the Brillouin zone have been done.
The parent contribution, however, requires evaluation at the part of the calculation
where eigenvalues and vectors are obtained,
which makes its calculation  time consuming.

There are four types of contributions to $d\psi/d\vec{\tau}$:
\begin{eqnarray}
\label{dp0}
        - \frac{d }{ d\vec{\tau}} \psi_{i}(\vec{k},\vec{r})
&=& 
	\text{i} \Bigl(
	\vec{\delta}^{(1)}_{\tau}
	+
	\vec{\delta}^{(2)}_{\tau}
	+
	\vec{\delta}^{(3)}_{\tau}
	+
	\vec{\delta}^{(4)}_{\tau} 
	\Bigr) \psi_i(\vec{k},\vec{r})
\end{eqnarray}
\begin{eqnarray}
\label{dp1}
	\vec{\delta}^{(1)}_{\tau} \psi_i(\vec{k},\vec{r})
&\equiv& 
	\Theta(\vec{r} \in {\cal{I}})
	\delta(\tau_i,\tau)\ 
	\hat{\vec{p}} \psi_i(\vec{k},\vec{r})
\end{eqnarray}
\begin{eqnarray}
\label{dp2}
	\vec{\delta}^{(2)}_{\tau} \psi_i(\vec{k},\vec{r})
&\equiv& 
	\delta(\tau_i,\tau)
	\sum_{\tau\Pr L}
	\Theta(s_{t\Pr}\sminus r_{\tau\Pr})
	\RU_{t\Pr L}(e_i,\vec{r}_{\tau\Pr})
	\Omega_{t\Pr\ell}(e_i,\kappa_i)
\nonumber\\
&&\hskip 60 pt \times\ 
	\, \left(\begin{matrix} 
                0 \\ 
		- \text{i}
		\vec{\nabla}_\tau B_{L,L_i} (\kappa_{i},\vec{\tau}\Pr\sminus\vec{\tau}_i,\vec{k}) \\
		 \end{matrix}\right)
\end{eqnarray}
\begin{eqnarray}
\label{dp3}
	\vec{\delta}^{(3)}_{\tau} \psi_i(\vec{k},\vec{r})
&\equiv& 
	\Theta(s_{t}\sminus r_{\tau})
	\sum_{L}
	\hat{\vec{p}}
	\RU_{tL}(e_i,\vec{r}_{\tau})
	\Omega_{t \ell}(e_i,\kappa_i)
	\RS_{L,L_{i}}(\kappa_{i}, \vec{\tau}\sminus\vec{\tau}_i, \vec{k})
\end{eqnarray}
\begin{eqnarray}
\label{dp4}
	\vec{\delta}^{(4)}_{\tau} \psi_i(\vec{k},\vec{r})
&\equiv& 
	-
	\Theta(s_{t}\sminus r_{\tau})
	\sum_{L}
	\RU_{tL}(e_i, \vec{r}_{\tau})
	\Omega_{t\ell}(e_i,\kappa_i)
\nonumber \\
&&\hskip 60 pt \times\ 
	\left(\begin{matrix} 
                0 \\ 
		- \text{i}
		\vec{\nabla}_\tau B_{L,L_{i}} (\kappa_{i},\vec{\tau}\sminus\vec{\tau}_i,\vec{k}) \\
		 \end{matrix}\right)
\end{eqnarray}
where $\hat{\vec{p}}$ is the momentum operator $-\text{i}\vec{\nabla}$.
The first two terms, Equations (\ref{dp1}) and (\ref{dp2}), are parent terms, 
changes in a basis due to a change in the site the basis is centered on.
The first term, Equation (\ref{dp1}), is the derivative of the wave function
in the interstitial region (Equation (\ref{intbasis0}) with respect to its parent site.
Since the gradient of a solution to the Helmholtz equation is a solution to the 
Helmholtz equation, matrix elements 
$\braket{\psi_{i}}{\hat{\vec{p}}\psi_{j}}_{\I}$
and
$\bra{\psi_{i}} -\nabla^{2} \ket{\hat{\vec{p}}\psi_{j}}_{\I}$
are calculated as integrals over the surface of the muffin-tin spheres. 
As in Equation (\ref{inov1}), when interstitial region tail parameters are the same,
the evaluation requires $\kappa^{2}$ derivatives of structure functions.
Working out this contribution proceeds as in Equation (\ref{inov1}), 
although arriving at a finite form requires identities such as
\begin{eqnarray}
        \sum_{\mu} \hat{\vec{e}}_{\mu}{\cal U}_{\tau_b}
        &\Bigl(&
                B_{\ell_a m_a, \ell_b\sminus1 \, m_b\sminus\mu}
                        (\kappa_b, \vec{\tau}_a\sminus\vec{\tau}_b, \vec{k})
                \gauntcof {\ell_b\sminus1} {m_b\sminus\mu} {\ell_b} {m_b} {1} {\mu}
                \ \kappa_b^2
\nonumber\\
                &-&
                B_{\ell_a m_a, \ell_b\splus1 \, m_b\sminus\mu}
                        (\kappa_b, \vec{\tau}_a\sminus\vec{\tau}_b, \vec{k})
                \gauntcof {\ell_b\splus1} {m_b\sminus\mu} {\ell_b} {m_b} {1} {\mu}
        \Bigr)
\nonumber\\
         = \sum_{\mu} \hat{\vec{e}}_{\mu}{\cal U}_{\tau_a}
        &\Bigl(&
                B_{\ell_a\splus1 \, m_a\splus\mu, \ell_b m_b}
                        (\kappa_b, \vec{\tau}_a\sminus\vec{\tau}_b, \vec{k})
                \gauntcof {\ell_a} {m_a} {\ell_a\splus1} {m_a\splus\mu} {1} {\mu}
\nonumber\\
                &-&
                B_{\ell_a\sminus1 \, m_a\splus\mu, \ell_b m_b}
                        (\kappa_b, \vec{\tau}_a\sminus\vec{\tau}_b, \vec{k})
                \gauntcof {\ell_a} {m_a} {\ell_a\sminus1} {m_a\splus\mu} {1} {\mu}
                ~\kappa_b^2
        \Bigr)
\nonumber\\
&&
\end{eqnarray}
Potential matrix elements $\bra{\psi_{i}}V\ket{\psi_{j}}$ are calculated using
Fourier series as in Sect. \ref{inmats} with gradients taken as discussed after
equation (\ref{radder1}).

\begin{figure}
\includegraphics[width=.6\textwidth]{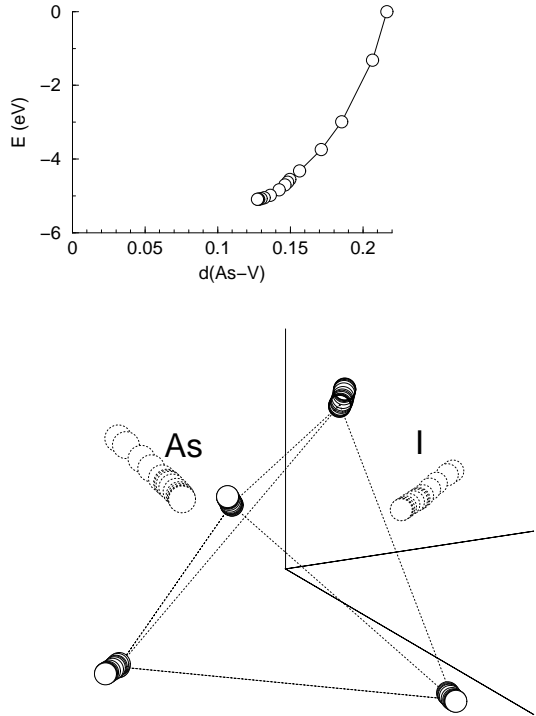}
\caption{Relaxation of a silicon 65 atom supercell containing a vacancy, 
a Si interstitial, and an As interstitial.
Of the 106 internal coordinates in this cell, 104 were allowed to relax
(2 coordinates were fixed to fix the center of mass of the crystal).
The calculation used a simple Broyden's method to zero atomic forces.}
\label{as_v_i_t}
\end{figure}

The second term, equation (\ref{dp0}), is the analog of the first term in the muffin-tin spheres;
{\it i.e.}, this term is the derivative of a basis with respect to its parent site 
evaluated in the muffin-tin spheres.
This term requires the gradient with respect to atomic positions of the structure function $B$.
This gradient is easily obtained from the structure function itself:
\begin{eqnarray}
      \vec{B}\Pr_{\ell m,\ell\Pr m\Pr}(\kappa,\vec{\tau}\sminus\vec{\tau}\Pr,\vec{k})
&\equiv&
      \frac{\partial }{ \partial \vec{u}}
      B_{\ell m, \ell\Pr m\Pr}(\kappa,\vec{u},\vec{k})
      \Big|_{\vec{u} = \vec{\tau}-\vec{\tau}\Pr}
\nonumber\\
&\equiv&
      \sum_{\mu} \text{i}\hat{\vec{e}}_{\mu} {\cal U}_{\tau}
      B^{\prime(\mu)}_{\ell m,\ell\Pr m\Pr}
            (\kappa,\vec{\tau}\sminus\vec{\tau}\Pr,\vec{k})
\nonumber\\
      B^{\prime(\mu)}_{\ell m,\ell\Pr m\Pr}
            (\kappa,\vec{\tau}\sminus\vec{\tau}\Pr,\vec{k})
&=&
      \Bigl(\ \ 
            \gauntcof {\ell}{m}{\ell\splus1}{m\splus\mu} 1 \mu
            B_{\ell\splus1 m\splus\mu, \ell\Pr m\Pr}
                  (\kappa,\vec{\tau}\sminus\vec{\tau}\Pr,\vec{k})
\nonumber\\
&& 
            -
            \kappa^2
            \gauntcof {\ell}{m}{\ell\sminus1}{m\splus\mu} 1 \mu
            B_{\ell\sminus1 m\splus\mu, \ell\Pr m\Pr}
                  (\kappa,\vec{\tau}\sminus\vec{\tau}\Pr,\vec{k})
      \Bigr)
\nonumber\\
&& \vec{\tau}-\vec{\tau}\Pr \neq 0
\label{sfdir} 
\end{eqnarray}
If 
convergence with respect to $\ell$ on the left hand side of the structure function
is sufficient for the energy, terms in $\ell_{{\rm max}}+1$ in Equation (\ref{sfdir})
may be neglected in evaluating forces.
As stated above, the evaluation of these terms is somewhat time consuming.

Examples of the use of forces for structural relaxation are given in Figures \ref{dist_parms} and 
\ref{as_v_i_t}.
Figure \ref{dist_parms} shows deviations from ideal lattice positions calculated for 
rhombohedral BaTiO$_{3}$ as a function of volume compared to experiment
\cite{kwei}. 
The rhombohedral angle was also relaxed at each volume in this calculation.
The Ti coordinate is a displacement along $[111]$. 
The oxygen displacements $\Delta x$ are along face diagonals while 
$\Delta z$ is toward the cell center.
These calculations included Ti 3$s$ and 3$p$ and Ba
5$s$ and 5$p$ along with the usual valence bases in a single, fully hybridizing basis.
At convergence, forces on internal coordinates were less than 1 mRy/Bohr.
Figure \ref{as_v_i_t} is a calculation of structural relaxation of As-vacancy-interstitial 
complex in Si. 
To a sixty-four atom Si supercell was added an As impurity at a tetrahedral interstitial position
and
a Si interstitial at an exchange position both surrounding a vacancy.
The crystal, far from equilibrium, was then allowed to relax.
Two internal coordinates (of a total of 106) were fixed to fix the center of mass 
of the crystal.
The energy was minimized with respect to the other 104 internal coordinates by 
zeroing the forces (to with 1 mRy/Bohr).
The forces were zeroed using a simple Broyden's method.


\section{Conclusion} 
In this article we have described our highly accurate 
full-potential LMTO method for solving the Kohn-Sham equations. 
In particular, we have shown that by dividing the crystal space 
into non-overlapping ``muffin-tin''  spheres and an interstitial region,
we can compute the charge density or the potential without any shape 
approximation, thus eliminating any need for empty spheres 
which are necessary in other LMTO implementations when the crystal is not 
closely packed. 
Another feature  of our implementation  
is that we can describe multiple principle quantum numbers
within  a single, {\it fully hybridized} basis set.
This is accomplished simply by using functions $\phi$ and $\dot\phi$
calculated with energies $\lbrace e_{n\ell}\rbrace$
corresponding to different principal quantum numbers $n$
to describe the radial dependence of a basis in the muffin-tin
region.
In the interstitial region our method uses ``multiple $\kappa$'' basis sets,
for a better description of the interstitial charge density. 
Highly accurate charge density can be obtained by systematically increasing 
 the number of variational parameters $\kappa$ for each
angular momentum of the basis set. 

The potential in a muffin-tin sphere at $\vec\tau$
has an expansion in linear combinations of spherical harmonics
invariant under
that part of the point group that leaves atomic positions invariant.
The evaluation of the interstitial potential matrix
only requires a correct treatment of basis functions (and potential) in
the interstitial region.
We have used this degree of freedom to design 
``pseudo-basis functions'', equal to 
the true basis functions  in the interstitial region  and are smooth
functions  in  the muffin-tin region, with the requirement that  
their Fourier transforms converge rapidly enough for practical use.

The set of internal forces acting on the atomic sites of a crystal
is a symmetric, discrete function of atom coordinates and
has a spherical expansion on the crystal sites with the same coefficients as
continuous symmetric functions. 
The total force on a site is given by the  derivative of the LDA total 
energy with respect to the atomic  position. 
Our implementation of the forces is
in many ways similar to that of Yu {\it et al.} for the LAPW method \cite{Yu0}. 
Because our basis set is a site-centered one, we are required to compute
additional terms, which can be  time consuming. These contributions to the
forces are non existant in plane-wave based methods, such as the 
pseudo-potential method.
In addition to   the ``Helmann-Feynman'' term, 
which accounts for the explicit dependence of the energy functional
A
on atomic  positions, the other contributions are: 
(1) an ``Incomplete Basis Set'' term, 
(2) a core-correction term, (3) a surface term arising from the change in
integration boundaries when atoms are moved, (4) a term which arises when the
calculation isn't fully self-consistent, and (5) a term arising from the
way in which the Brillouin zone integrals are performed. 
We have showed that the forces are accurate enough 
to relax atomic structures. 
As examples, forces have been used to optimize the
internal coordinates of rhombohedral BaTiO$_{3}$ as a function of volume
and the geometry of a 65 atom As, vacancy, and interstitial defected Si 
supercell.
Where experimental results are available, good agreement 
is obtained.

\section{Acknowledgments} 

One of us J.M.W would like to thank  
the Universit\'e Louis Pasteur 
  for  his IPCMS stay. M.A and O.E  
 collaboration is partially supported by the TMR network 'Interface
Magnetism' of the European Commission (Contract No. EMRX-CT96-0089) .

\clearpage
\addcontentsline{toc}{section}{Index}
\flushbottom
\printindex
\end{document}